# Overcoming the Efficiency–Stability Trade-off in Spin–Orbit Torque Devices with Thermally Robust BCC NiW Alloys

**Yu-Ming Pan†[1], Chen-Yi Wei†[1], Yi-Cheng Tsou[1], Tsung-Yu Pan, Guang-Yu Guo[2], Chih-Huang Lai[1*]**

[1]Department of Materials Science and Engineering, National Tsing Hua University, Hsinchu 300044, Taiwan

[2]Department of Physics, National Taiwan University, Taipei 10617, Taiwan

# Abstract

The development of high-performance spin–orbit torque (SOT) magnetic memories is fundamentally constrained by a persistent trade-off between spin Hall efficiency, thermal structural stability, and perpendicular magnetic anisotropy in conventional heavy metals. Here, we overcome this limitation by engineering body-centered-cubic (BCC) Ni-doped W alloys as highly efficient and thermally robust spin-current sources. $Ni_{30}W_{70}$/CoFeB heterostructures achieve deterministic out-of-plane magnetization switching at an ultra-low critical current density of 1.78 MA/cm², nearly threefold lower than that of β-W, while maintaining a high anisotropy field of 8,500 Oe and a thermal stability factor of 57.9. The BCC $Ni_{30}W_{70}$ alloy preserves its structural integrity and the perpendicular magnetic anisotropy of the adjacent CoFeB layer after annealing at 450 °C, demonstrating robustness under the stringent thermal processing conditions relevant to back-end-of-line integration. Harmonic Hall and ferromagnetic resonance measurements reveal a large spin Hall angle of −0.39 and a high interfacial spin transparency of 0.75, demonstrating efficient spin-current generation and interfacial transmission. First-principles calculations further reveal enhanced intrinsic spin Hall conductivity in W-rich BCC NiW alloys, associated with the Fermi level lying within a spin–orbit-coupling-induced band gap. These findings establish BCC NiW alloys as a scalable and thermally resilient material platform for energy-efficient SOT-MRAM.

†These authors contributed equally: Yu-Ming Pan, Chen-Yi Wei.

*Correspondence and requests for materials should be addressed to Chih-Huang Lai: chlai@mx.nthu.edu.tw

# Introduction

Spin–orbit torque magnetic random-access memory (SOT-MRAM) has emerged as a promising platform for energy-efficient, high-speed, and non-volatile memory owing to its separate write and read pathways and potential compatibility with advanced semiconductor technologies. In conventional SOT heterostructures, a charge current flowing through a nonmagnetic spin-source layer generates a transverse spin current through the spin Hall effect (SHE), which is transmitted across the interface and exerts a torque on the magnetization of an adjacent ferromagnetic (FM) layer [1,2]. Efficient SOT switching therefore requires both efficient spin-current generation within the spin-source material and its transmission across the spin-source/FM interface, while preserving the magnetic stability required for non-volatile operation.

Heavy metals with strong spin–orbit coupling, including Ta, Pt, and W, have been extensively investigated as SOT sources [3-7]. Among them, metastable β-W has become an important benchmark owing to its large spin Hall angle (SHA) and high electrical resistivity [3,4,8]. Alloying and doping provide additional routes to manipulate the spin Hall response through changes in electronic structure, resistivity, and scattering, as demonstrated in W–Si, Pt–Cr, Pt–Cu, and Au–W systems [9-13]. Within W-based materials, Ta alloying and nitrogen incorporation have also been employed to enhance or preserve the spin Hall response[14-17].

First-principles studies further demonstrate that compositional modification of W can shift the Fermi level relative to spin–orbit-coupling-induced band anti-crossings and band gaps, thereby substantially modulating the intrinsic spin Hall conductivity.[14,18]

A persistent challenge, however, is to achieve high SOT efficiency without sacrificing structural and magnetic stability. The high-efficiency β-W phase and many of its derivatives are metastable and can transform toward the thermodynamically stable BCC α-W phase upon high-temperature annealing, accompanied by substantial degradation of the spin Hall response[15-20]. Such structural evolution can also affect the spin-source/FM interface and perpendicular magnetic anisotropy (PMA) of the adjacent FM layer. These limitations become particularly critical for back-end-of-line (BEOL) integration, where thermal processing approaching 400 °C imposes stringent requirements on structural and interfacial stability. Alternative approaches, including nitrogen-stabilized W[17], topological insulators[21], and ordered intermetallic compounds such as $Ni_4W$[22], offer promising spin-transport characteristics, but challenges associated with thermal robustness, materials integration, epitaxial growth, or switching efficiency remain. An effective SOT material must therefore reconcile efficient spin-current generation and transmission with robust magnetic and high-temperature structural stability.

Ni–W alloys provide an intriguing yet comparatively unexplored materials space for addressing this efficiency–stability trade-off. Ni-rich NiW has previously been employed as a functional non-ferromagnetic underlayer in magnetic multilayers, where FCC NiW(111) promotes the heteroepitaxial growth of subsequent layers, demonstrating its compatibility with technologically relevant magnetic heterostructures[23]. Related Ni-based refractory-metal alloys exhibit substantial and composition-tunable electrical resistivity[24], motivating exploration of Ni alloying for resistivity engineering because increased resistivity can enhance the SHA when a large intrinsic spin Hall conductivity is retained. Whereas previous studies have primarily exploited Ni-rich FCC NiW for crystallographic texture control, the complementary W-rich BCC regime remains largely unexplored for spin-current generation. Despite extensive theoretical investigations of dopant and impurity effects in W[14,25], the influence of Ni incorporation on the intrinsic SHE of W-rich BCC NiW also remains unclear. These considerations motivate exploration of W-rich $Ni_xW_{1-x}$ as a materials space for simultaneously tuning electrical resistivity and intrinsic spin Hall response while pursuing the structural and interfacial stability required for efficient SOT switching.

Here, we demonstrate that W-rich BCC NiW alloys overcome this persistent efficiency–stability trade-off. $Ni_{30}W_{70}$/CoFeB heterostructures achieve deterministic out-of-plane magnetization switching at an ultra-low critical current density of 1.78 $MA/cm^2$, nearly threefold lower than the β-W benchmark, while retaining robust PMA in the adjacent CoFeB layer and structural integrity after annealing at 450 °C. Harmonic Hall and ferromagnetic resonance measurements reveal a large spin Hall angle of −0.39 and an interfacial spin transparency of 0.75, revealing efficient spin-current generation and interfacial transmission. First-principles calculations further reveal an enhanced intrinsic spin Hall conductivity in W-rich BCC NiW alloys, which is associated with the Fermi level being positioned within a spin–orbit-coupling-induced band gap. These results reveal that the exceptional switching performance arises from the interplay between efficient bulk spin-current generation and interfacial spin transmission, while the stable BCC structure provides the high-temperature resilience required for semiconductor processing. BCC NiW thus provides a promising materials platform for energy-efficient SOT devices compatible with the back-end-of-line (BEOL) thermal processing.

# Results and discussion

### Composition-dependent spin–orbit torque in $Ni_xW_{1-x}$ alloys

To identify the optimal Ni–W composition for efficient spin-current generation, we systematically investigated Ti(2)/$Ni_xW_{1-x}$(6)/Co(2)/Ti(4) heterostructures with x ranging from 0.1 to 0.9 (Fig. 1a). The Co layer exhibits an in-plane magnetic easy axis along the y-direction for all compositions, as confirmed by vibrating sample magnetometry (Supplementary Note 1), allowing the composition-dependent current-induced switching behavior to be compared without substantial variation in the underlying magnetic anisotropy. The observed evolution of the switching characteristics can therefore be primarily associated with changes in the structural and spin-transport properties of the $Ni_xW_{1-x}$ layer.

X-ray diffraction reveals a pronounced evolution of the NiW structure with composition (Fig. 1b). In the W-rich regime, the diffraction feature associated with BCC W (110) progressively broadens as the Ni concentration increases, indicating reduced long-range crystalline order. At higher Ni concentrations, the Ni (111)-related reflection becomes increasingly prominent. No additional diffraction peaks attributable to secondary intermetallic phases are observed, indicating that the sputtered NiW layers remain compositionally mixed over the investigated range. This structural evolution is accompanied by a substantial change in electrical resistivity. As shown in Fig. 1c, the resistivity ρ increases strongly upon alloying and reaches a maximum near $Ni_{30}W_{70}$, where ρ is approximately 244 μΩ cm.

We next examined how this compositional evolution influences current-induced magnetization switching. Figure 1d shows representative Y-type SOT switching loops for W-rich and Ni-rich alloys. A clear reversal of the switching polarity is observed across the composition series. Ni-rich alloys ($x \geq 0.7$) exhibit a positive switching polarity, whereas increasing the W fraction reverses the switching polarity, with a compensation region occurring near $x \approx 0.5$. The negative switching polarity observed in the W-rich compositions is consistent with the negative spin Hall response characteristic of W-based spin-current sources.[4,8]

The critical switching current density, Jc, was determined by accounting for current partitioning among the conducting layers using their independently measured resistivities, as detailed in Supplementary Note 2. A pronounced non-monotonic dependence of Jc on Ni concentration is observed (Fig. 1f). Pure α-W and β-W exhibit Jc values of approximately 14.3 and 5.7 $MA/cm^2$, respectively, whereas $Ni_{30}W_{70}$ reaches a substantially reduced Jc of approximately 1.8 $MA/cm^2$ (Fig. 1e). Thus, Ni incorporation does not simply interpolate between the spin-transport properties of elemental Ni and W; instead, an optimum emerges within the W-rich compositional regime.

Plotting Jc against resistivity provides further insight into this behavior (Fig. 1g). Across much of the composition series, an increase in resistivity is accompanied by a reduction in switching current. Such a correlation is consistent with previous observations that resistivity engineering can enhance the effective spin Hall angle and SOT efficiency when a substantial spin Hall conductivity is retained.[9,10,26] Notably, $Ni_{30}W_{70}$ deviates from a simple resistivity-based interpretation: despite having a resistivity comparable to high-resistivity β-W, it exhibits a substantially lower switching-current density. This deviation indicates that resistivity enhancement alone is insufficient to account for the exceptional SOT efficiency of $Ni_{30}W_{70}$, suggesting additional contributions from its intrinsic spin Hall response and spin-current transmission across the NiW/ferromagnet interface. These contributions are quantified in the following sections through harmonic Hall, ferromagnetic resonance, interfacial structural analyses, and first-principles calculations.

The composition-dependent structural, electrical, and switching measurements identify $Ni_{30}W_{70}$ as the optimal composition within the experimentally investigated series, combining high resistivity, the lowest switching-current density, and a W-like negative SOT polarity. We therefore selected $Ni_{30}W_{70}$ for integration with CoFeB/MgO perpendicular-anisotropy heterostructures and subsequent evaluation of its structural stability, spin-transport properties, and out-of-plane switching performance.

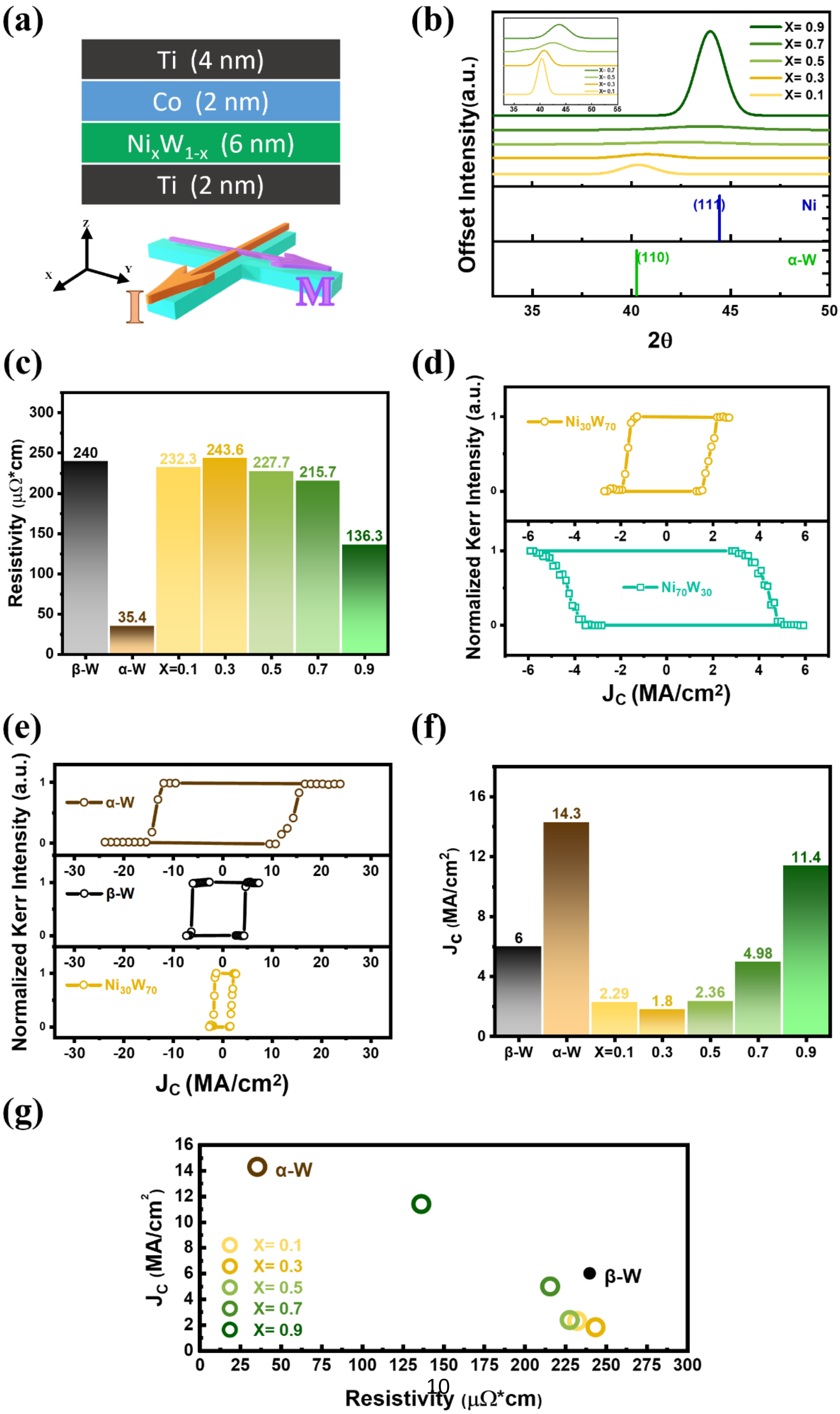
(a)
Ti (4 nm)
Co (2 nm)
Ni$_x$W$_{1-x}$ (6 nm)
Ti (2 nm)
I
M
(b)
X= 0.9
X= 0.7
X= 0.5
X= 0.3
X= 0.1
(111)
Ni
(110)
α-W
Offset Intensity(a.u.)
2θ
(c)
240
35.4
232.3
243.6
227.7
215.7
136.3
Resistivity (μΩ*cm)
β-W
α-W
X=0.1
0.3
0.5
0.7
0.9
(d)
Ni$_{30}$W$_{70}$
Ni$_{70}$W$_{30}$
Normalized Kerr Intensity (a.u.)
J$_C$ (MA/cm$^2$)
(e)
α-W
β-W
Ni$_{30}$W$_{70}$
Normalized Kerr Intensity (a.u.)
J$_C$ (MA/cm$^2$)
(f)
6
14.3
2.29
1.8
2.36
4.98
11.4
J$_C$ (MA/cm$^2$)
β-W
α-W
X=0.1
0.3
0.5
0.7
0.9
(g)
α-W
β-W
X= 0.1
X= 0.3
X= 0.5
X= 0.7
X= 0.9
J$_C$ (MA/cm$^2$)
Resistivity (μΩ*cm)

**Fig. 1 | Composition-dependent structural, electrical, and spin–orbit torque properties of NiW alloys.**

**a,** Schematic of the Ti(2)/$Ni_xW_{1-x}$ (6)/Co(2)/Ti(4) heterostructure and measurement geometry for in-plane SOT switching, with the Co magnetization oriented along the y-axis and the pulsed current applied along the x-axis. Layer thicknesses are given in nanometres. **b,** X-ray diffraction (XRD) patterns of $Ni_xW_{1-x}$ films with different Ni concentrations, together with reference diffraction peaks for Ni and α-W. **c,** Electrical resistivity as a function of Ni concentration x, including α-W and β-W reference films. **d,** Representative current-induced in-plane SOT switching loops for W-rich $Ni_{30}W_{70}$ and Ni-rich $Ni_{70}W_{30}$, showing the reversal of switching polarity with composition. **e,** Comparison of in-plane SOT switching loops for α-W, β-W, and $Ni_{30}W_{70}$. **f,** Critical switching-current density $J_c$ as a function of Ni concentration x, including the α-W and β-W references. **g,** Relationship between $J_c$ and electrical resistivity for the investigated $Ni_xW_{1-x}$ compositions and W reference films.

## Thermally robust BCC NiW/CoFeB heterostructures

Having established efficient composition-dependent SOT switching in $Ni_xW_{1-x}$, we next investigated whether the optimized $Ni_{30}W_{70}$ alloy could be integrated into a perpendicular magnetic anisotropy (PMA) heterostructure relevant to high-density SOT-MRAM. This transition is technologically important because perpendicular CoFeB/MgO-based magnetic structures provide the magnetic stability and scalability required for non-volatile memory, while placing stringent demands on the structural and interfacial integrity of the underlying spin-current source during high-temperature processing. In particular, simultaneously preserving efficient SOT generation, robust PMA, and structural stability under the thermal budget required for semiconductor integration remains a central materials challenge. We therefore fabricated Ta(2)/$Ni_{30}W_{70}$(6)/$Co_{20}Fe_{60}B_{20}$(1)/MgO(2)/Ta(4) heterostructures and subjected them to post-deposition annealing at 450 °C for 1 h, providing a stringent test of whether the $Ni_{30}W_{70}$ spin-source layer can withstand the thermal processing required for device integration. $Ni_{30}W_{70}$ was selected because it exhibited the lowest critical switching-current density among the compositions investigated in Fig. 1.

The structural and chemical integrity of the annealed heterostructure was examined by cross-sectional transmission electron microscopy (TEM) and energy-dispersive X-ray spectroscopy (EDS). The cross-sectional TEM image in Fig. 2a reveals a continuous multilayer structure with well-defined interfaces after annealing at 450 °C. Elemental mapping (Fig. 2c) and the corresponding compositional depth profile (Fig. 2b) show spatially uniform distributions of Ni and W within the NiW layer, with no discernible Ni enrichment at the NiW/CoFeB interface. Moreover, no appreciable diffusion of the bottom Ta layer into the NiW/CoFeB region is observed. These results demonstrate that the $Ni_{30}W_{70}$ layer maintains its chemical integrity and well-defined interfaces after the 450 °C thermal treatment.

To determine whether the NiW layer also retains its desired crystal structure after high-temperature annealing, we performed site-specific nano-beam electron diffraction (NBD) using an approximately 5-nm electron probe positioned within the $Ni_{30}W_{70}$ layer (Fig. 2a). The resulting diffraction pattern exhibits distinct reflections consistent with a BCC-like structure. Analysis of the diffraction pattern yields an estimated lattice parameter of approximately 2.93 Å, smaller than that of elemental BCC W, consistent with lattice contraction upon incorporation of the smaller Ni atoms into the W-rich lattice. Together with the composition-dependent XRD results in Fig. 1, these local structural measurements demonstrate that W-rich $Ni_{30}W_{70}$ retains a contracted BCC-like structure after annealing at 450 °C.

The simultaneous preservation of the W-rich BCC-like structure, homogeneous Ni–W distribution, and well-defined NiW/CoFeB interface distinguishes $Ni_{30}W_{70}$ from metastable β-W-based spin sources, for which high-temperature processing can induce structural transformation and degradation of spin-transport properties[15,16,19,20]. The structural resilience of $Ni_{30}W_{70}$ at 450 °C therefore establishes the materials foundation for combining high-temperature process compatibility with efficient SOT generation. We next examine whether this structural robustness is accompanied by the magnetic stability and low-current perpendicular switching required for SOT-MRAM operation.

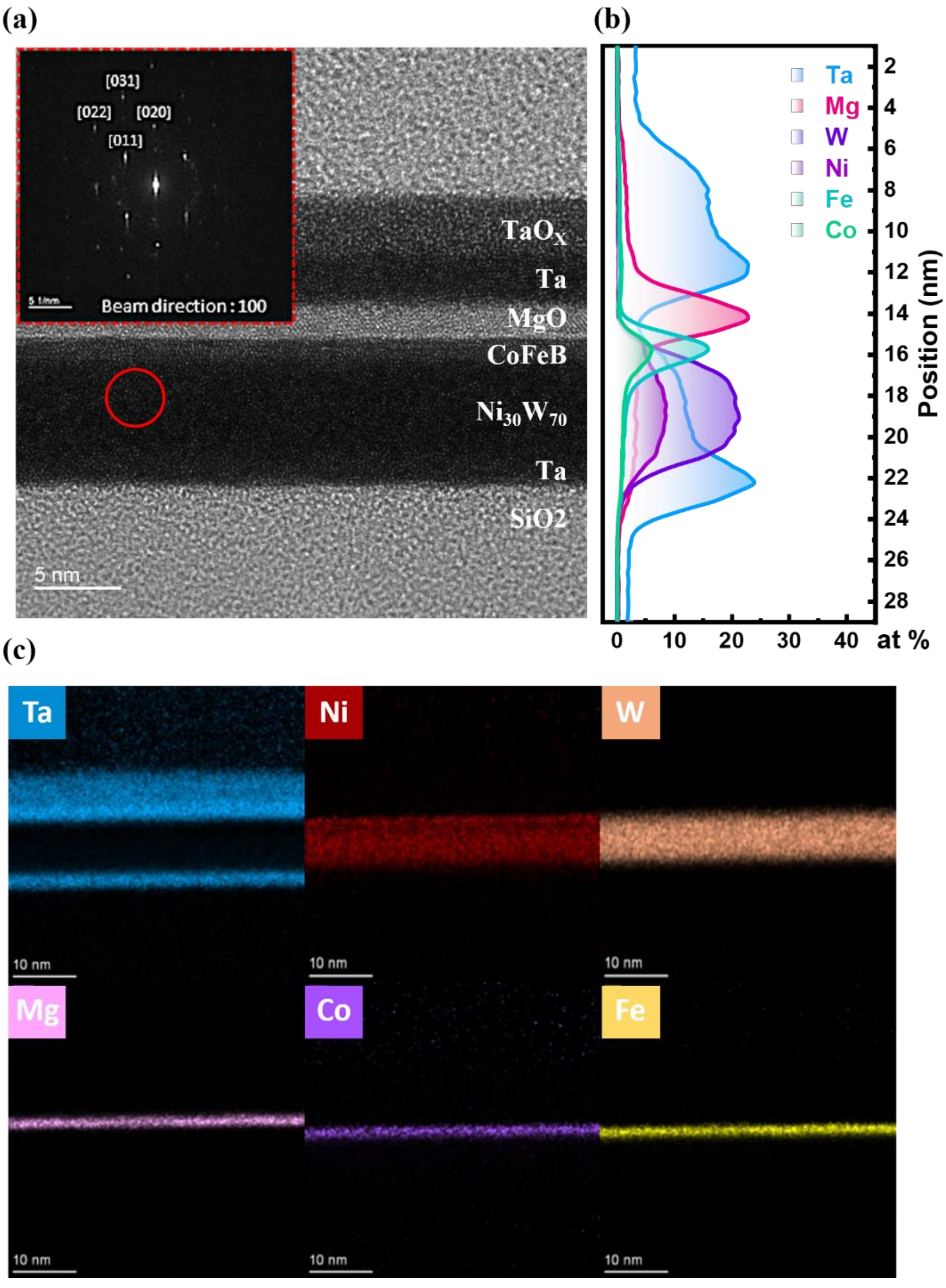

**Fig. 2 | Structural and chemical stability of the $Ni_{30}W_{70}$/CoFeB heterostructure after 450 °C annealing.**

**a,** Cross-sectional transmission electron microscopy (TEM) image of the Ta(2)/$Ni_{30}W_{70}$ (6)/CoFeB(1)/MgO(2)/Ta(4) heterostructure after annealing at 450 °C for 1 h. The inset shows the site-specific nano-beam electron diffraction (NBD) pattern acquired from the $Ni_{30}W_{70}$ layer at the position indicated by the red circle, consistent with a BCC-like structure. Layer thicknesses are given in nanometres. **b,** EDS compositional depth profile across the multilayer stack, showing the elemental distributions through the heterostructure. **c,** Cross-sectional EDS elemental maps of Ta, Ni, W, Mg, Co and Fe, showing the spatial distribution of the constituent elements after annealing.

## Low-current SOT switching with robust magnetic stability

Having established the structural integrity of the $Ni_{30}W_{70}$/CoFeB heterostructure after 450 °C annealing, we next examined whether this thermal resilience is accompanied by the magnetic stability and switching efficiency required for perpendicular SOT devices. Figure 3a shows the room-temperature in-plane and out-of-plane magnetic hysteresis loops of the annealed Ta/$Ni_{30}W_{70}$/CoFeB/MgO/Ta heterostructure. The pronounced difference between the two field orientations confirms robust perpendicular magnetic anisotropy (PMA), with an anisotropy field $H_k$ of approximately 8500 Oe. This retention of strong PMA after 450 °C processing contrasts with metastable β-W-based heterostructures, in which high-temperature annealing can lead to structural transformation and interfacial degradation[15,27,28]. Consistent with the preserved interfacial integrity observed in Fig. 2, thickness-dependent magnetic measurements yield a magnetic dead-layer thickness of only 0.33 nm (Supplementary Note 3), compared with values up to approximately 0.53 nm reported for W/CoFeB interfaces under high-temperature processing[15,27,28]. These results demonstrate that the $Ni_{30}W_{70}$ underlayer supports robust PMA in the adjacent CoFeB layer even after the 450 °C thermal treatment.

We next evaluated current-induced perpendicular magnetization switching. Figure 3b shows the anomalous Hall resistance as a function of applied current under different in-plane assist fields $H_x$, using a current-pulse width of 15 μs. Reversal of the switching polarity with the direction of $H_x$ confirms the SOT-driven nature of the magnetization reversal. At $H_x$=500 Oe, corresponding to only approximately 5.9% of $H_k$, the $Ni_{30}W_{70}$/CoFeB device exhibits a critical switching-current density $J_c$=1.78 MA/cm$^2$. This value is nearly threefold lower than that obtained for the β-W/CoFeB reference system, demonstrating that the low switching current identified in the in-plane composition screening is retained upon integration into a perpendicular CoFeB/MgO heterostructure.

The simultaneous realization of a large anisotropy field and low switching-current density is further reflected by the $H_k/J_c$ figure of merit, which reaches approximately 4.8 Oe·cm²/kA for $Ni_{30}W_{70}$/CoFeB, more than an order of magnitude larger than the ~0.45 Oe·cm²/kA reported for conventional W/CoFeB heterostructures.[29] This comparison highlights that the reduced switching current in NiW is achieved without compromising the magnetic anisotropy required for stable perpendicular magnetization.

To further quantify the thermal stability of the perpendicular magnetic state, we examined the pulse-width dependence of SOT switching using circular devices with a diameter of 5 μm (Fig. 3c). The critical switching current was measured as a function of pulse width τ and analyzed using a thermally assisted switching model,[30]

$$I_c = I_{c0}\left[1 - \frac{1}{\Delta}\ln\left(\frac{\tau_{pulse}}{\tau_0}\right)\right] \#(1)$$

where $I_{c0}$ is the zero-thermal-fluctuation critical current, Δ is the thermal stability factor, and $\tau_0$ is the intrinsic attempt time, taken to be 1 ns[10]. Fitting the pulse-width dependence in Fig. 3d yields a zero-thermal critical current density $J_{c0}$=10.2 MA/cm$^2$ and a thermal stability factor Δ=57.9. This value lies within the range generally targeted for long-term non-volatile data retention.[31] Moreover, $J_{c0}$ remains substantially lower than representative values reported for β-W/CoFeB (~32 MA/cm$^2$)[4] and Ta-based heterostructures (~40 MA/cm$^2$)[3].

These results establish that the low switching current of $Ni_{30}W_{70}$ is not achieved at the expense of magnetic stability. Instead, the 450 °C-annealed heterostructure simultaneously combines Jc=1.78 MA/cm$^2$, Hk≈8500 Oe, and Δ=57.9, demonstrating that efficient SOT switching can coexist with robust perpendicular anisotropy and thermal stability in a W-rich BCC NiW spin-source platform.

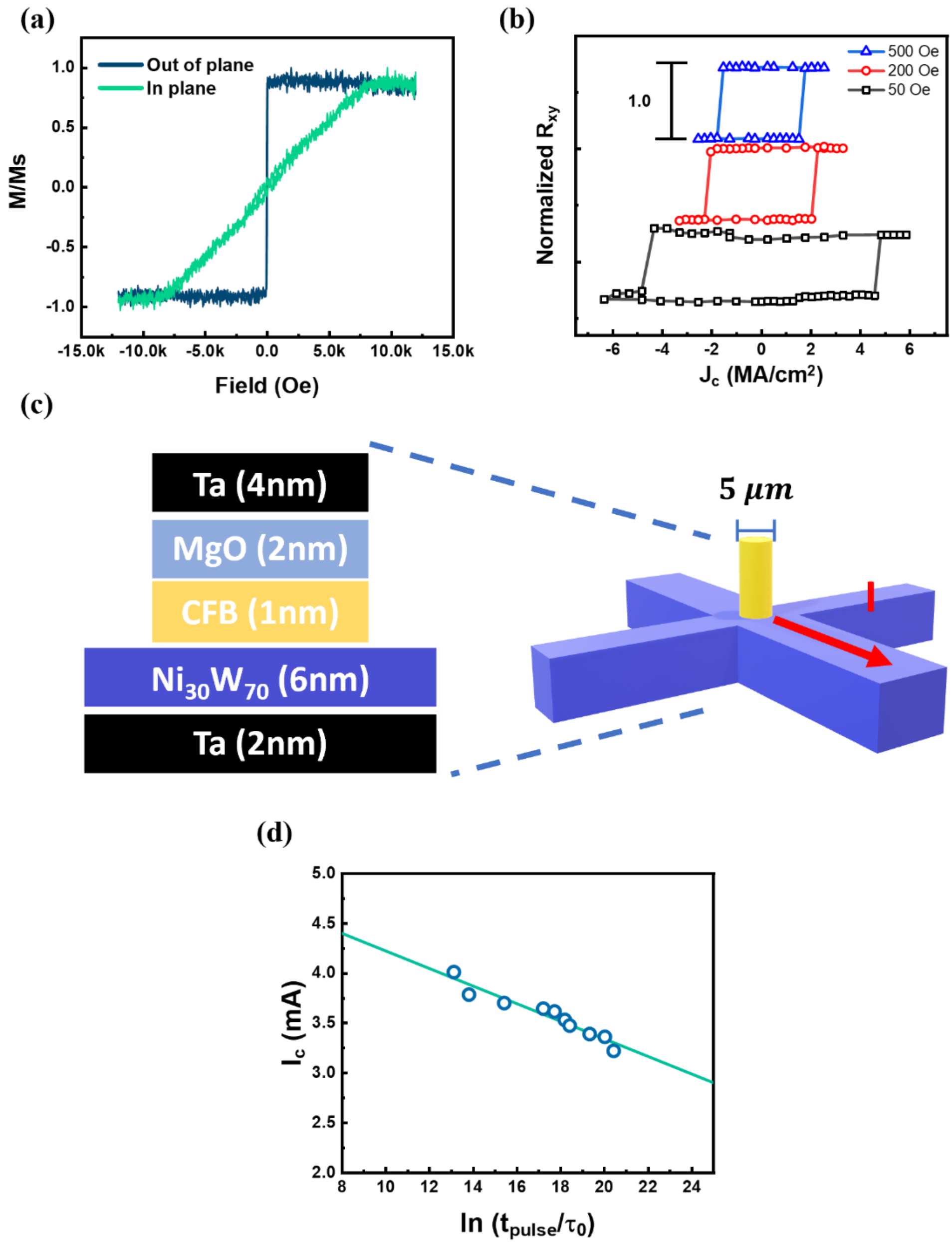

**Fig. 3 | Perpendicular magnetic anisotropy, SOT switching and thermal stability of $Ni_{30}W_{70}$/CoFeB heterostructures.**

**a,** In-plane and out-of-plane magnetic hysteresis loops of the annealed Ta/$Ni_{30}W_{70}$/CoFeB/MgO/Ta heterostructure, showing perpendicular magnetic anisotropy with an anisotropy field $H_k$ of approximately 8500 Oe. **b,** Current-induced out-of-plane (Z-type)

SOT switching loops measured under different in-plane assist fields $H_x$ using a fixed current-pulse width of $\tau = 15\ \mu s$. **c,** Schematic of the circular Ta/$Ni_{30}W_{70}$/CoFeB/MgO/Ta device used for pulse-width-dependent switching measurements. **d,** Critical switching current as a function of current-pulse width under $H_x = 500\ Oe$. The solid line represents the fit to the thermally assisted switching model used to extract the zero-thermal-fluctuation critical current and thermal stability factor $\Delta$ .

**Efficient spin-current generation and interfacial transmission**

To elucidate the origin of the low switching-current density in $Ni_{30}W_{70}$, we first examined the dependence of $J_c$ on the $Ni_{30}W_{70}$ thickness $t_{NiW}$. As shown in Fig. 4a, $J_c$ exhibits a non-monotonic thickness dependence, initially decreasing with increasing $t_{NiW}$, reaching a minimum near 6 nm, and subsequently increasing in the thicker regime ( $t_{NiW}$>6 nm). The initial decrease is consistent with the progressive accumulation of spin current generated within the NiW layer as its thickness approaches and exceeds the characteristic spin-diffusion length. At larger thicknesses, additional NiW contributes increasingly to charge-current conduction without a proportional increase in the spin current reaching the NiW/CoFeB interface, resulting in a higher total current required for switching[3,10]. This thickness dependence therefore indicates that efficient switching requires an appropriate balance between spin-current generation and charge-current distribution within the NiW layer. To quantify the charge-to-spin conversion efficiency, we performed harmonic Hall measurements as a function of $Ni_{30}W_{70}$ thickness ($t_{NiW}$), as detailed in Supplementary Note 4. The effective spin Hall angle $\theta_{SH}^{eff}$ extracted from the harmonic Hall measurements increases in magnitude with increasing $Ni_{30}W_{70}$ thickness and approaches saturation above approximately 6 nm (Fig. 4b). The thickness dependence was fitted using the spin-diffusion model [8,32,33].

$$\theta_{SH}^{eff}(t_{NiW}) = \theta_{SH}^{\infty}\left[1 - sech\left(\frac{t_{NiW}}{\lambda}\right)\right] \tag{2}$$

where $\theta_{SH}^{\infty}$ denotes the bulk-limit spin Hall angle, and λ is the spin-diffusion length of $Ni_{30}W_{70}$. The fit yields $\theta_{SH}^{\infty}$=−0.39 and λ=2.5 nm. The magnitude of $\theta_{SH}^{\infty}$ is comparable to that reported for high-efficiency β-W (∼−0.40), whereas the extracted spin-diffusion length is shorter than the ∼3.5 nm reported for β-W[32]. These measurements establish that W-rich $Ni_{30}W_{70}$ retains a large negative spin Hall response despite adopting the thermally robust BCC-like structure identified in Fig. 2.

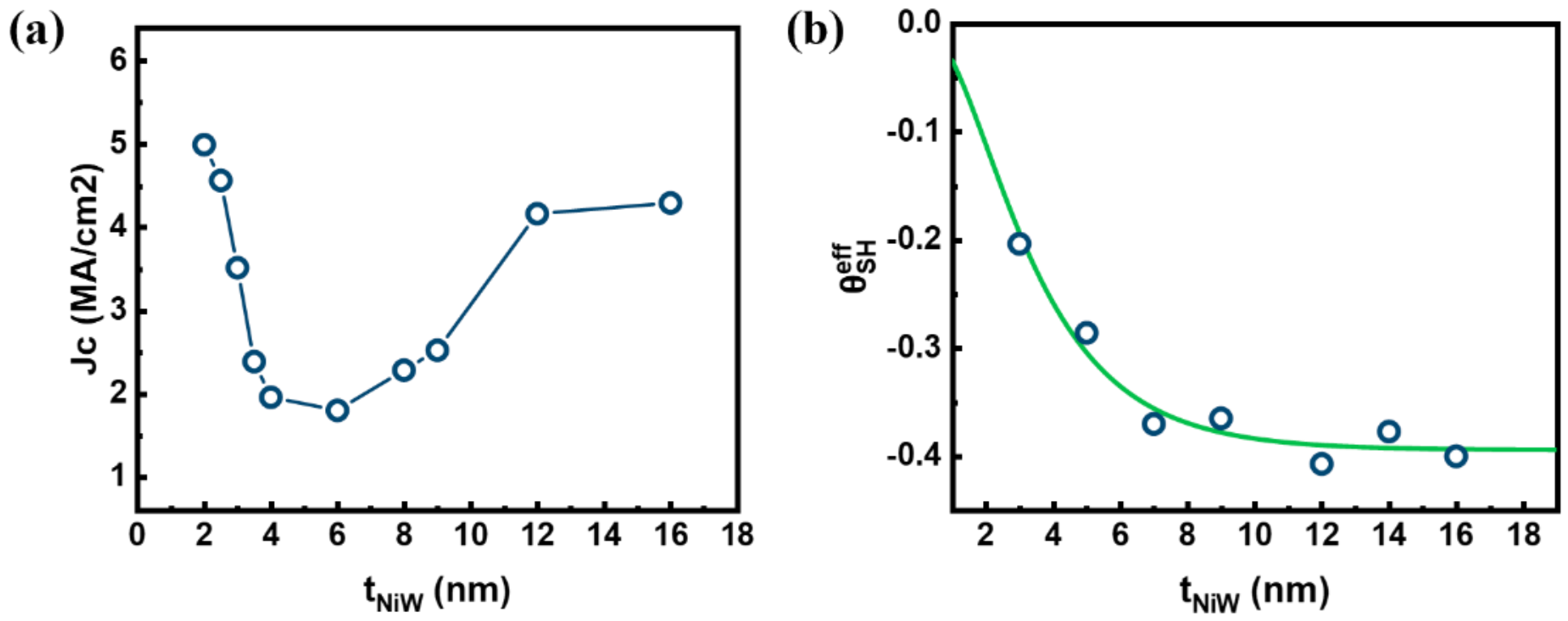

**Fig. 4 | Thickness dependence of SOT switching and spin Hall response in NiW.**

**a,** Critical switching-current density $J_c$ as a function of $Ni_{30}W_{70}$ thickness $t_{NiW}$ . **b,** Effective spin Hall angle $\theta_{SH}^{eff}$ as a function of $t_{NiW}$. The solid line represents a fit to the spin-diffusion model, yielding a bulk-limit spin Hall angle $\theta_{SH}^{\infty}$=−0.39 and a spin diffusion length λ=2.5 nm.

A large spin Hall angle alone, however, does not fully account for the exceptionally low $J_c$ of $Ni_{30}W_{70}$. The spin Hall angle describes the efficiency of spin-current generation within the bulk spin-source layer, whereas the torque exerted on the CoFeB magnetization additionally depends on the fraction of the generated spin current transmitted across the NiW/CoFeB interface. We therefore quantified the interfacial spin transparency T to distinguish these bulk and interfacial contributions.

To quantify the efficiency of spin-current transmission across the $Ni_{30}W_{70}$/CoFeB interface, we evaluated the interfacial spin transparency within the spin-diffusion framework developed for spin Hall magnetoresistance, in which spin transport in the nonmagnetic layer is coupled to the interface through the spin-mixing conductance.[34] The interfacial spin transparency is expressed as

$$T = \frac{G_{\uparrow\downarrow} tanh\left(\frac{d}{2\lambda}\right)}{G_{\uparrow\downarrow} coth\left(\frac{d}{\lambda}\right) + \frac{\sigma_{NM}}{\lambda}\frac{h}{2e^2}} \#(3)$$

where $d$, $\lambda$ and $\sigma_{NM}$ are the layer thickness, spin diffusion length, and electrical conductivity of the $Ni_{30}W_{70}$ layer, respectively, and $G_{\uparrow\downarrow}$ is the spin-mixing conductance of the NiW/CoFeB interface. The imaginary component of $G_{\uparrow\downarrow}$ is assumed to be negligible compared with its real component[35]. The interfacial spin-mixing conductance $G_{\uparrow\downarrow}$ is related to the experimentally accessible effective spin-mixing conductance $G_{eff}^{\uparrow\downarrow}$, which includes spin-backflow effects, through[34]:

$$G_{\uparrow\downarrow} = G_{\uparrow\downarrow}^{eff} \frac{\frac{\sigma_{NM}}{\lambda}\frac{h}{2e^2}}{\frac{\sigma_{NM}}{\lambda}\frac{h}{2e^2} - G_{\uparrow\downarrow}^{eff}} \#(4)$$

where $G_{\uparrow\downarrow}^{eff}$ is given by:

$$G_{\uparrow\downarrow}^{eff} = \frac{4\pi M_S t_F}{g\mu_B}\left(\alpha_{NM/FM} - \alpha_{FM}\right) \#(5)$$

Here, $M_s$ is the saturation magnetization of CoFeB, $t_F$ is the ferromagnetic layer CoFeB thickness, g is the electron g-factor, and $\mu_B$ is the Bohr magneton. The damping parameters of $Ni_{30}W_{70}$ (6 nm)/CoFeB (t)/MgO heterostructures, $\alpha_{NM/FM}$, and corresponding CoFeB reference structures without the $Ni_{30}W_{70}$ underlayer, $\alpha_{FM}$, were experimentally determined using strip-line ferromagnetic resonance measurements, as detailed in Supplementary Note 5. As shown in Fig. 5, the damping enhancement, $(\alpha_{NM/FM} - \alpha_{FM})$ $M_s$, scales linearly with $1/t_F$, allowing $G_{\uparrow\downarrow}^{eff}$ to be extracted from the slope.

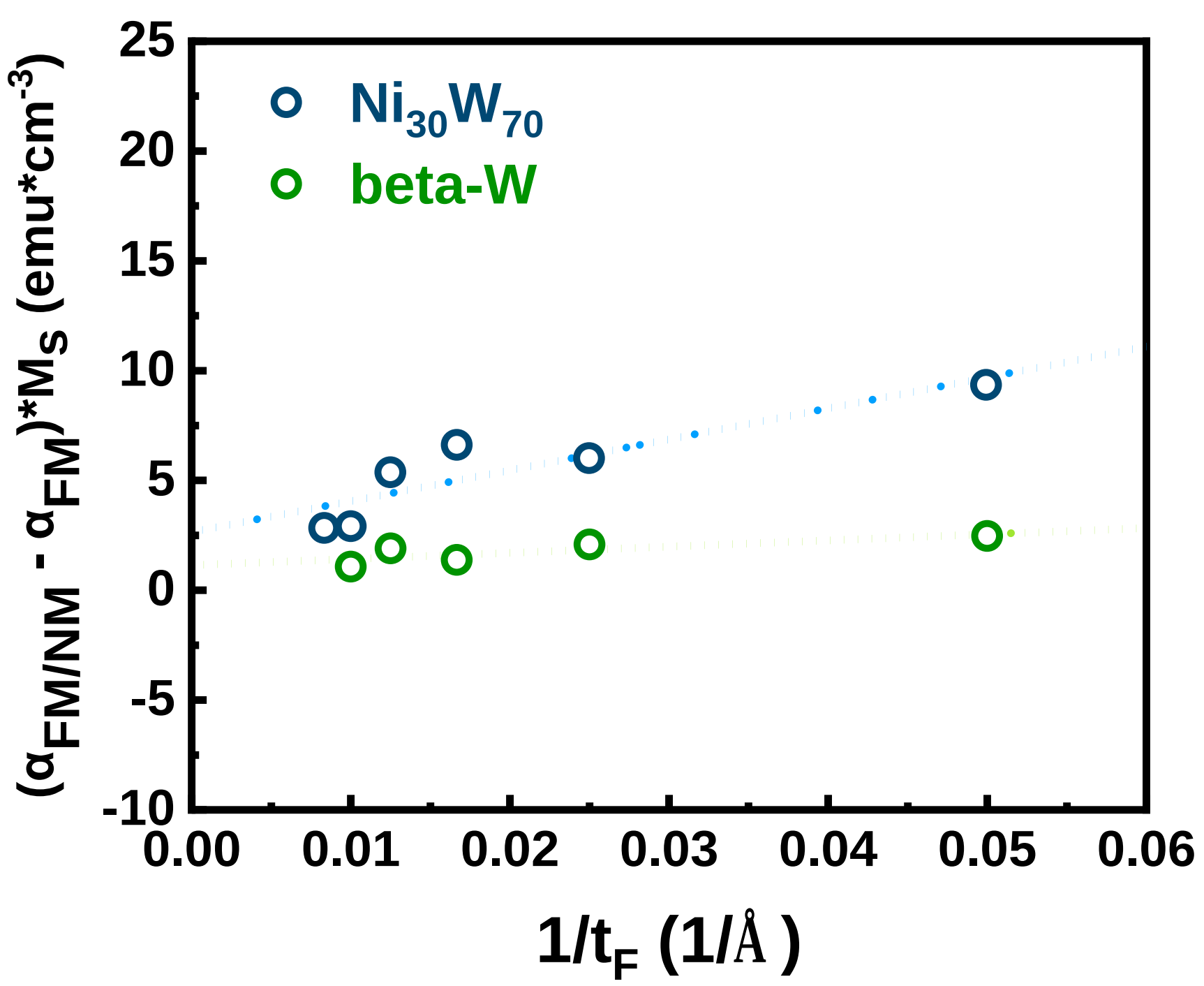

**Fig. 5 | Determination of the effective spin-mixing conductance.** $(\alpha_{NM/FM} - \alpha_{FM})$Ms as a function of inverse CoFeB thickness, $1/t_F$, for $Ni_{30}W_{70}$/CoFeB and β-W/CoFeB heterostructures. Dashed lines represent linear fits used to extract $G_{\uparrow\downarrow}^{eff}$.

The resulting effective spin-mixing conductance, $G_{\uparrow\downarrow}^{eff}$ , for $Ni_{30}W_{70}$/CoFeB is $9.53\times10^{18}$ $m^{-2}$, approximately five times that obtained for the β-W/CoFeB reference ($1.91\times10^{18}$ $m^{-2}$). Using the experimentally determined resistivity and spin-diffusion length, we obtain an interfacial transparency T=0.75 for $Ni_{30}W_{70}$/CoFeB, compared with T=0.66 for β-W/CoFeB. The $Ni_{30}W_{70}$/CoFeB value is among the higher interfacial transparencies reported for W-based spin-source systems[36]. These results demonstrate that the advantage of $Ni_{30}W_{70}$ arises not only from its large bulk spin Hall response but also from efficient transmission of the generated spin current across the $Ni_{30}W_{70}$ /CoFeB interface.

To examine whether the enhanced spin transmission correlates with the structural quality of the interface, we further compared $Ni_{30}W_{70}$ /CoFeB and β-W/CoFeB heterostructures using X-ray reflectivity (XRR). The measurements were performed across the same CoFeB-thickness series used for the FMR analysis, enabling direct comparison between interfacial structure and spin-transport properties. The extracted root-mean-square interfacial roughness, as shown in Fig. 6, remains consistently low for $Ni_{30}W_{70}$ /CoFeB, with an average value of approximately 0.28 nm, compared with approximately 0.71 nm for β-W/CoFeB. The substantially smoother NiW/CoFeB interface is consistent with the larger $G_{\uparrow\downarrow}^{eff}$ and higher T, supporting more efficient interfacial spin transmission. Together with the small magnetic dead layer and chemically well-defined interface retained after 450 °C annealing (Figs. 2 and 3), these results highlight the pronounced interfacial robustness of the $Ni_{30}W_{70}$/CoFeB heterostructure. This behavior contrasts with previously reported W-based alloy systems, in which high-temperature processing can induce substantial microstructural disorder and intermixing at the spin-source/FM interface[15]. Collectively, the structural and spin-transport measurements establish that $Ni_{30}W_{70}$ combines efficient bulk spin-current generation with a robust interface that enables efficient spin-angular-momentum transfer to CoFeB. The exceptionally low switching current of $Ni_{30}W_{70}$ therefore cannot be attributed to a single materials parameter. Rather, it emerges from the combined contributions of a large bulk-limit spin Hall angle ($\theta_{SH}^{\infty}$=−0.39) and high interfacial spin transparency (T=0.75), while the thermally robust BCC-like structure preserves the structural and magnetic integrity required for device operation. The electronic origin of the enhanced intrinsic spin Hall conductivity in W-rich BCC NiW is examined below using first-principles calculations.

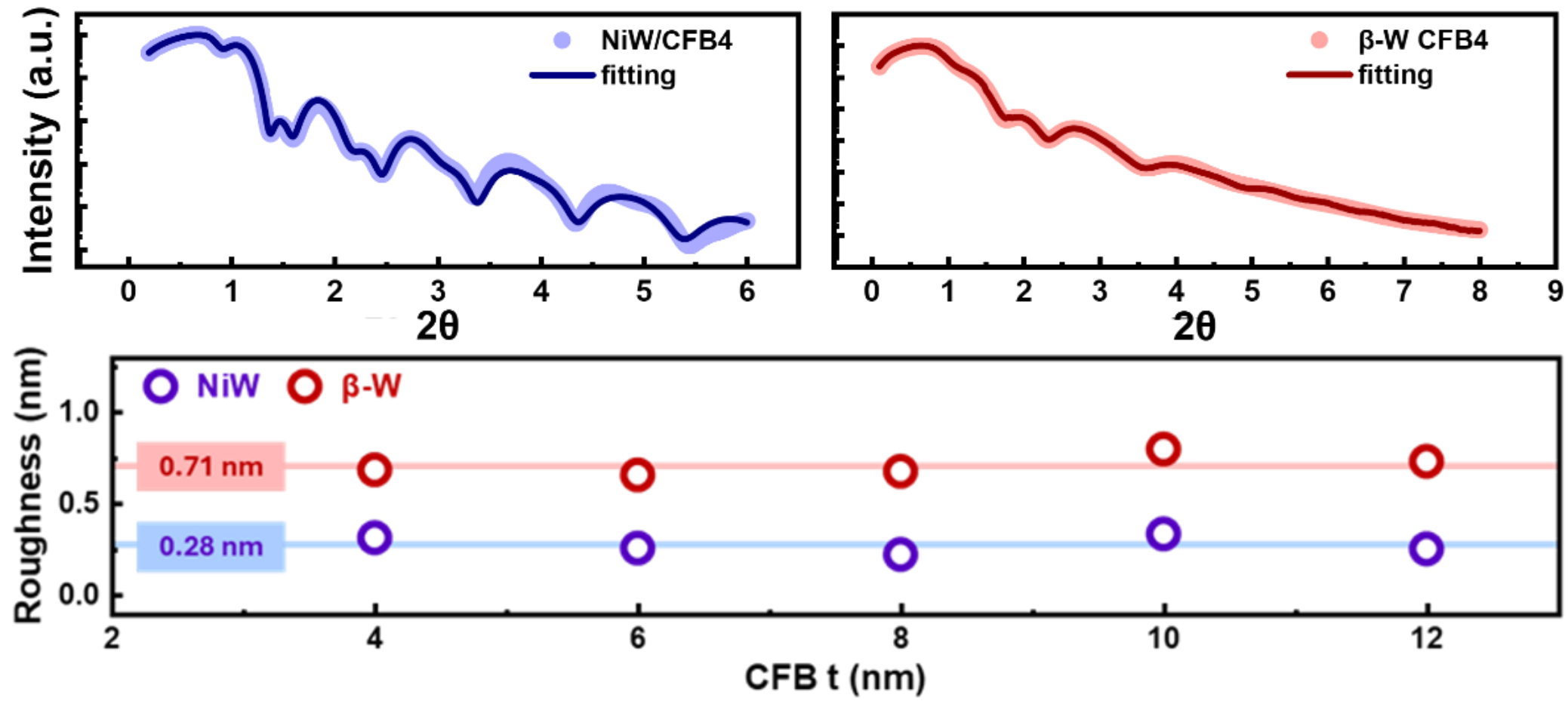

**Fig. 6 | Interfacial structural quality of $Ni_{30}W_{70}$/CoFeB and β-W/CoFeB heterostructures.** Representative X-ray reflectivity (XRR) spectra and corresponding fits for $Ni_{30}W_{70}$ (6 nm)/CoFeB (4 nm) and β-W(6 nm)/CoFeB (4 nm) heterostructures. The lower panel shows the extracted root-mean-square interfacial roughness as a function of CoFeB thickness, yielding average values of approximately 0.28 nm and 0.71 nm for $Ni_{30}W_{70}$/CoFeB and β-W/CoFeB, respectively.

**Electronic origin of enhanced intrinsic spin Hall conductivity in W-rich BCC NiW**

To elucidate the electronic origin of the enhanced spin Hall response in W-rich NiW, we performed first-principles calculations of the electronic structures and intrinsic spin Hall conductivity (SHC) of BCC $Ni_xW_{1-x}$ alloys and pure BCC W (Supplementary Note 6). The intrinsic SHC was evaluated from the spin Berry curvature using the Kubo formalism. The calculations explicitly considered substitutional Ni concentrations of x=0.0625, 0.125, and 0.25 in BCC W, together with pure W; an idealized x=0.5 computational model was also examined for comparison.

As summarized in Table 1, pure BCC W exhibits a calculated SHC of -$485(\hbar/e)(\Omega\,\mathrm{cm})^{-1}$. Ni incorporation substantially enhances the magnitude of the intrinsic SHC in the W-rich regime, yielding $-575$, $-870$, and $-660$ $(\hbar/e)(\Omega\,\mathrm{cm})^{-1}$ for Ni concentrations of 6.25, 12.5, and 25 at%, respectively. Importantly, the calculated SHC remains negative throughout these W-rich compositions, consistent with the experimentally determined sign of the spin Hall response. When combined with the measured resistivities, the corresponding estimated spin Hall angles are also substantially enhanced relative to pure BCC W (Table 1). Although metastable β-W is known to exhibit a large intrinsic SHC associated with its characteristic A15 electronic structure, the present calculations address a different question: whether Ni incorporation can enhance the intrinsic spin Hall response within the thermally robust BCC W framework. Previous first-principles calculations indeed predict a larger intrinsic SHC for β-W than for BCC α-W, consistent with its experimentally large spin Hall angle[18]. Our results show that substantial enhancement of the intrinsic SHC can also be achieved within the W-rich BCC NiW phase, without relying on the metastable A15 structure.

**Table 1 |** Calculated spin Hall conductivity ($\sigma_{SH}$), together with the measured resistivity (ρ) of pure α-W and its $Ni_xW_{1-x}$ alloys. The corresponding spin Hall angles ($\theta_{SH}$) are also listed, which are given by $\theta_{SH} = \left(\frac{2e}{\hbar}\right)\sigma_{SH}\rho$.

| | **α-W** | **$Ni_{1/16}W_{15/16}$** | **$Ni_{2/16}W_{14/16}$** | **$Ni_{4/16}W_{12/16}$** | **$Ni_{1/2}W_{1/2}$** |
|---|---|---|---|---|---|
| **$\sigma_{SH}$ (ħ/e)(Ω-cm)$^{-1}$** | **-485** | **-575** | **-870** | **-660** | **-288** |
| **ρ (μΩ-cm)** | **35.4** | **232.3** | **232.3** | **243.6** | **227.7** |
| **$\theta_{SH}$** | **-0.034** | **-0.267** | **-0.404** | **-0.322** | **-0.131** |

*The resistivity value $\rho$ is estimated from the nearest experimentally measured composition.

To identify the microscopic origin of this enhancement, we further examined the relativistic electronic band structure and density of states of $Ni_{4/16}W_{12/16}$ as a representative W-rich BCC $Ni_xW_{1-x}$ alloy (Supplementary Fig. S8). Comparison of the band structures calculated with and without spin–orbit coupling (SOC) reveals that SOC lifts near-degeneracies along the X–M symmetry line and opens a local gap encompassing the Fermi level $E_F$. Such SOC-gapped near-degeneracies can generate pronounced spin Berry-curvature contributions from the SOC-split bands. When contributions of opposite sign from both branches are present, they tend to cancel; when $E_F$ lies within the SOC-induced gap, however, this compensation is reduced, resulting in an enhanced contribution to the intrinsic SHC[37]. This interpretation is consistent with established intrinsic-SHE physics, in which SOC-gapped crossings close to $E_F$ can generate large spin Berry curvature. Such SOC-induced gaps at near-degenerate band crossings are known to generate large Berry-curvature contributions to the intrinsic spin Hall conductivity in heavy metals and A15 compounds[14,36]. Our calculations therefore indicate that the enhanced intrinsic SHC in W-rich BCC NiW is associated with SOC-induced reconstruction of the electronic bands near $E_F$, providing a microscopic electronic-structure basis for the enhanced spin Hall response upon Ni alloying.

Importantly, the composition dependence predicted by the first-principles calculations should not be interpreted as a direct prediction of the experimentally optimized switching composition. Among the explicitly modeled W-rich alloys, the largest calculated intrinsic SHC occurs at 12.5 at.% Ni, whereas the lowest experimental switching current is obtained near $Ni_{30}W_{70}$ (Fig. 1). This difference reflects the distinct quantities probed by theory and experiment. The calculations describe the intrinsic electronic contribution to the SHC of idealized BCC NiW structures, whereas the experimentally measured switching current is governed by the combined effects of spin-current generation, electrical resistivity, spin diffusion, magnetic properties, and spin-current transmission across the NiW/CoFeB interface.

This distinction is particularly relevant for $Ni_{30}W_{70}$, for which our experiments reveal a large bulk-limit spin Hall angle of $\theta_{SH}^{\infty}=-0.39$, an interfacial spin transparency of T=0.75, and a substantially enhanced effective spin-mixing conductance. The first-principles calculations therefore identify an intrinsic electronic mechanism by which Ni incorporation enhances the spin Hall conductivity of W-rich BCC NiW, while the experimentally optimized switching at $Ni_{30}W_{70}$ emerges from the combined contributions of bulk spin-current generation and efficient interfacial spin transmission. This complementary picture explains why the composition maximizing the calculated intrinsic SHC need not coincide with that minimizing the experimentally measured switching current.

## Discussion

To place W-rich BCC NiW in the broader landscape of SOT materials, we benchmark the critical switching-current density (Jc), electrical resistivity (ρ), thermal stability factor (Δ), and anisotropy field ($H_k$) against representative heavy-metal and alloy-based SOT systems (Table 2). $Ni_{30}W_{70}$/CoFeB exhibits a Jc of 1.78 MA/cm$^2$, among the lowest values reported for representative β-W/CoFeB and W-alloy-based systems under their respective measurement conditions, while retaining a moderate resistivity of approximately 244 μΩ cm. Importantly, this low switching current is accompanied by Δ=57.9, $H_k$=8500 Oe, and retention of perpendicular magnetic anisotropy and structural integrity after annealing at 450 °C. NiW therefore occupies a distinct performance regime in which low switching current, magnetic stability, and high-temperature structural robustness are achieved concurrently rather than optimized individually.

Alloying and chemical modification have been widely employed to enhance the spin Hall response of heavy metals through changes in electronic structure, resistivity, and scattering[12,17,19,38]. In several W-based systems, however, enhanced SOT efficiency is accompanied by increased resistivity or relies on metastable structures whose spin-transport and interfacial properties can deteriorate during high-temperature processing[15,17,27]. These considerations are particularly relevant for device integration because reducing switching current alone does not necessarily minimize write-energy dissipation or ensure compatibility with the thermal budget of semiconductor processing. The significance of $Ni_{30}W_{70}$ therefore lies not in maximizing any single spin-transport parameter, but in reconciling efficient switching with electrical transport, magnetic stability, and thermal process robustness within the same material system.

More broadly, our results point to a materials-design principle for SOT heterostructures that extends beyond maximizing the bulk spin Hall response alone. The preceding transport, FMR, structural, and first-principles analyses collectively show that efficient device-level switching depends on the entire pathway from spin-current generation in the bulk, through spin-angular-momentum transmission across the interface, to preservation of the magnetic and structural properties during processing. W-rich NiW provides an example in which these requirements can be jointly engineered within a thermally robust BCC-based alloy. This perspective shifts the optimization of W-based SOT materials from maximizing an isolated conversion metric toward co-designing bulk spin generation, interfacial spin transmission, and thermal stability. Further studies in nanoscale magnetic tunnel junctions and field-free switching architectures will be required to determine how these materials-level advantages translate into integrated SOT-MRAM devices.

**Table 2 |** Comparison of the electrical resistivity, thermal stability factor Δ, critical switching-current density $J_c$, anisotropy field $H_k$ and annealing temperature of $Ni_{30}W_{70}$ /CoFeB with representative SOT material systems reported in the literature.

| System | Resistivity ($\mu\Omega * cm$) | Thermal stability (Δ) | Jc (MA/cm2) ($H_x$/pulse width) | $H_k$ (Oe) | Annealing Temperature | Ref. |
|---|---|---|---|---|---|---|
| Cu-Pt/Co (PMA) | 82.5 | 28.85 | 2.37 ($H_x$=600 Oe/50ms) | - | - | 9 |
| Cr-Pt/Co (PMA) | 133 | 33.58 | 3.43 ($H_x$=1000 Oe/50ms) | 8000 | - | 10 |
| V-Pt/Co (PMA) | 83 | 45.45 | 5.29 ($H_x$=2600 Oe/50ms) | - | - | |
| Si-W/CFB (PMA) | 135 | 50.84 | 17 ($H_x$=100 Oe/10µs) | - | 300/500℃ | 12 |
| N-W/CFB (PMA) | 500 | - | 5.7 ($H_x$=200 Oe/10µs) | 7000 | 300℃ | 16 |
| V-W/CFB (PMA) | 103-110 | - | 20 ($H_x$=50 Oe/-) | - | 300℃ | 39 |
| β-W/CFB (PMA) | - | - | 17 ($H_x$=100 Oe/-) | - | 250℃ | 4 |
| **$Ni_{30}W_{70}$/CFB** (PMA) | **244** | **58** | **1.78** **(**$H_x$=500 Oe/15µs**)** | **8500** | **450**℃ | **This work** |

"–" indicates that the corresponding value was not reported in the cited study.

# Methods

## Film growth and device fabrication

Multilayer thin films with nominal stack structures of Ti(2)/$Ni_xW_{1-x}$(6)/Co(2)/Ti(4) for in-plane magnetization studies and Ta(2)/$Ni_{30}W_{70}$(t)/$Co_{20}Fe_{60}B_{20}$(1)/MgO(2)/Ta(4) for perpendicular magnetic anisotropy (PMA) studies were deposited on thermally oxidized Si substrates with a 200-nm-thick $SiO_2$ layer using DC/RF magnetron sputtering. Numbers in parentheses denote nominal layer thicknesses in nanometers. The base pressure of the deposition chamber was maintained below $5\times10^{-8}$ Torr, and deposition was performed under an ultra-high-purity Ar working pressure of 3 mTorr.

The $Ni_xW_{1-x}$ alloy layers were deposited by co-sputtering from high-purity elemental Ni (99.99%) and W (99.95%) targets. The alloy composition was systematically varied by adjusting the respective sputtering powers of the Ni and W targets and was experimentally determined by energy-dispersive X-ray spectroscopy (EDS). The bottom Ti or Ta layer served as a buffer/adhesion layer, whereas the top Ti or Ta layer served as a protective capping layer.

All samples designed for PMA measurements were subjected to post-deposition annealing at 450 °C for 1 h under high vacuum ($<1\times10^{-7}$ Torr) without an applied magnetic field. Room-temperature magnetic hysteresis loops were characterized using a vibrating sample magnetometer (VSM) with magnetic fields applied both in and out of the film plane.

For electrical transport and SOT switching measurements, the multilayer films were patterned into Hall-cross devices with a channel width of 20 μm and a length of 80 μm using photolithography and ion-beam etching. Ta(5 nm)/Pt(100 nm) electrical contacts were subsequently fabricated by a lift-off process.

**Spin–orbit torque switching measurements**

For in-plane (Y-type) SOT switching measurements of Ti/$Ni_xW_{1-x}$/Co/Ti devices, the Co magnetization was initially saturated along the y-direction using an external magnetic field. After removal of the field, current-induced magnetization switching was monitored using the magneto-optical Kerr effect (MOKE), with the Kerr signal probing the in-plane magnetization component.

For out-of-plane (Z-type) SOT switching measurements of Ta/$Ni_{30}W_{70}$/CoFeB/MgO/Ta devices, the perpendicularly magnetized CoFeB layer was initially saturated along the z-direction. After removal of the perpendicular saturation field, current-induced switching was measured under an in-plane assist field $H_x$ applied along the current direction. Current pulses were supplied using a Keithley 6221 current source, and the magnetic state was read through the anomalous Hall effect (AHE) using a Keithley 2000 multimeter. The switching measurements were carried out with a current pulse width of 15 μs.

The critical switching current density, $J_c$, was determined by accounting for current partitioning among the conductive layers using their respective electrical resistivities, as detailed in Supplementary Note 2. For the thermal-stability analysis, circular devices with a diameter of 5 μm were fabricated, and the pulse-width dependence of the switching current was analyzed using a thermally assisted switching model. An intrinsic attempt time of 1 ns was used in the analysis.

**Harmonic Hall measurements**

Current-induced SOT effective fields and the spin Hall angle were evaluated using harmonic Hall measurements. A low-frequency AC current with amplitudes ranging from 1 to 12 mA and a frequency of 20.5 Hz was supplied to the Hall-bar devices using a Keithley 6221 current source. The first- and second-harmonic Hall voltages, ($V_{\omega}$) and ($V_{2\omega}$), were detected using a 7280BFP lock-in amplifier while sweeping the applied in-plane magnetic field. The harmonic Hall responses were analyzed to extract the current-induced effective fields and the effective spin Hall angle, as described in detail in Supplementary Note 4. The thickness dependence of the extracted spin Hall angle was subsequently fitted using a spin-diffusion model to determine the bulk-limit spin Hall angle and spin-diffusion length of $Ni_{30}W_{70}$.

## Ferromagnetic resonance measurements

Ferromagnetic resonance (FMR) measurements were performed using a strip-line technique to characterize magnetic damping and quantify spin transmission across the $Ni_{30}W_{70}$/CoFeB interface. A series of $Ni_{30}W_{70}$(6 nm)/CoFeB($t_{FM}$)/MgO samples with varying CoFeB thicknesses were measured together with corresponding CoFeB reference samples without the $Ni_{30}W_{70}$ underlayer. The damping enhancement induced by the adjacent $Ni_{30}W_{70}$ layer was determined from the difference between the damping parameters of the bilayer and reference samples. Detailed FMR procedures are provided in Supplementary Note 5.

The product $(\alpha_{NM/FM}-\alpha_{FM})\times$Ms was analyzed as a function of $1/t_{FM}$, and the effective spin-mixing conductance $G_{eff\uparrow\downarrow}$ was extracted from the linear thickness dependence. The interfacial spin transparency T was subsequently calculated using the spin-mixing conductance together with the experimentally determined $Ni_{30}W_{70}$ resistivity and spin-diffusion length, following the spin-pumping framework described in the main text. The same analysis was performed for β-W/CoFeB reference structures to provide a direct comparison.

## Structural and interfacial characterization

The crystallographic structures of the $Ni_xW_{1-x}$ alloy films were characterized by X-ray diffraction (XRD) using Cu $K_\alpha$ radiation. The composition-dependent diffraction patterns were used to evaluate the structural evolution from W-rich BCC NiW toward Ni-rich structures. Cross-sectional transmission electron microscopy (TEM), nano-beam diffraction (NBD), and energy-dispersive X-ray spectroscopy (EDS) were used to investigate the microstructure, crystallographic phase, elemental distribution, and interfacial integrity of the Ta/$Ni_{30}W_{70}$/CoFeB/MgO/Ta heterostructures after annealing at 450 °C for 1 h. NBD measurements were performed using a focused electron probe positioned within the $Ni_{30}W_{70}$ layer to determine its local crystal structure. The resulting diffraction pattern was used to identify the contracted BCC-like NiW structure.

X-ray reflectivity (XRR) measurements were performed on $Ni_{30}W_{70}$/CoFeB and β-W/CoFeB heterostructures to evaluate layer thicknesses and interfacial roughness. The XRR measurements were conducted on the same CoFeB-thickness series used for the FMR damping analysis, enabling direct correlation between interfacial structure and spin-transport properties. The complete thickness-dependent XRR data are provided in the Supplementary Information.

**First-principles calculations**

First-principles calculations were performed to investigate the electronic structure and intrinsic spin Hall conductivity of W-rich BCC $Ni_xW_{1-x}$ alloys. Computational methods, structural models, electronic-structure calculations, and evaluation of the intrinsic spin Hall conductivity are described in Supplementary Note 6.

# Data availability

The data that support the findings of this study are available within the Article and its Supplementary Information files.

## Acknowledgements

This work was supported by the National Science and Technology Council, Taiwan, under Grant NSTC 114-2223-E-007-006 and 115-2223-E-007-003. The authors would like to thank the National Synchrotron Radiation Research Center, Taiwan, for their technical assistance with X-ray Diffraction.

## Author contributions

Y.-M.P. and C.-Y.W. contributed equally to this work. Y.-M.P. and C.-Y.W. designed and performed the majority of the experiments and analyzed the experimental data. Y.-C.T. and T.-Y.P. contributed to thin-film deposition and spin–orbit torque measurements. G.-Y.G. performed the first-principles calculations and analyzed the theoretical results. C.-H.L. conceived and supervised the project. Y.-M.P., C.-Y.W. and C.-H.L. wrote the manuscript. All authors discussed the results and contributed to the preparation of the manuscript.

## Competing interests

The authors declare no competing interests.